\documentclass{iau}
\usepackage{graphicx,natbib}
\title[Dark matter inner slope and concentration in galaxies]{Dark matter inner slope and concentration in galaxies: from the Fornax
  dwarf to M87} 

\author[Mamon et al.]{G. A. Mamon$^1$, J. Chevalier$^1$, A. J.
  Romanowsky$^{2,3}$ \and R. Wojtak$^4$}

\affiliation{$^1$Institut d'Astrophysique de Paris, UMR 7095 (CNRS \& UPMC),
  98 bis Bd Arago, F-75014 Paris, France, email: {\tt gam@iap.fr,
    jill.chevalier@gmail.com}\\
$^2$Dept. of Physics and Astronomy, San Jos\`e State University, 
San Jos\`e, CA 95192, USA\\
$^3$Univ. of California Observatories, 
Santa Cruz, CA
 95064, USA, email: {\tt romanow@ucolick.org}\\
$^4$Dark Cosmology Centre, 
DK-2100 Copenhagen, Denmark, email: {\tt wojtak@stanford.edu}}

\pubyear{2014}
\volume{311}
\jname{Galaxy Masses as Constraints of Formation Models}
\editors{M. Cappellari \& S. Courteau, eds.}

\begin{document}

\maketitle

\begin{abstract}
We apply two new state-of-the-art methods that model the
distribution of observed tracers in projected phase space to 
lift the mass / velocity anisotropy (VA) degeneracy and deduce constraints on the mass
profiles
of galaxies, as well as their VA.
We first show how a distribution function based method
applied to the satellite kinematics of otherwise isolated SDSS galaxies shows
convincing observational evidence of age matching: red
galaxies have more concentrated dark matter (DM) halos than blue galaxies of the
same stellar or halo mass.
Then, applying the MAMPOSSt technique to M87 (traced by its red and blue
globular clusters) we find that very cuspy DM is favored, unless we release
priors on DM concentration or stellar mass (leading to unconstrained slope).
For the Fornax dwarf spheroidal  (traced by its
metal-rich and metal-poor stars), 
the inner DM slope is unconstrained, with weak evidence for a core if the
stellar mass is fixed. This highlights how priors are crucial for DM
modeling. Finally, we find that blue GCs around M87 and metal-rich stars in
Fornax have tangential outer VA.

\keywords{galaxies: elliptical and lenticular, cD - dark matter}
\end{abstract}

\firstsection
\section{Introduction}
\label{intro}

While DM cusps (with slopes steeper than $\gamma=-0.9$ for radii $r>r_{-2}/100$,
i.e. $r > r_{\rm vir}/1000$)
are predicted in DM-only cosmological $N$-body simulations
\citep{Navarro+04},  it is not clear how the astrophysics of
baryons will affect the slope $\gamma$ of the inner DM density profile. \cite{PG12} have shown
that intermittent feedback can transform the DM cusps into homogeneous cores,
while hydrodynamical cosmological $N$-body simulations by \cite{DiCintio+14}
lead to the view that $\gamma$ is a function of the
stellar to halo mass ratio, where $L^*$ ellipticals and dwarf spheroidals
(dSphs) are
expected to be cuspy while galaxies of intermediate mass should have DM cores.

Mass modeling of dSphs (using internal motions as the only possible method
available, and now with over 1000 and 200 member velocities available for
Sculptor and Fornax, respectively) leads to a variety of results.
For Sculptor, \cite{WP11} used the \cite{Wolf+10} pinch (insensitivity of the
mass profile to the a priori unknown velocity anisotropy [VA] at some ``pinch''
radius) on metal-rich ($Z$-rich) vs. metal-poor ($Z$-poor) stars to deduce a
shallow DM slope for 
Sculptor.
However, this result has been disputed by \cite{Breddels+13} and
\cite{RF14}, who both found (using orbit modeling and dispersion-kurtosis
Jeans modeling, respectively) that both cores and cusps were allowed for the DM.
While \cite{JG12}, using orbit modeling, deduced cored DM for the Fornax dSph,
\cite{AAE13} (using the Wolf pinch) found
that while a DM core is the natural outcome for Fornax,  (cuspy) NFW DM is possible if
its concentration ($r_{\rm vir}/r_{-2}$) is low. 

Similarly, two recent studies of the M87 giant elliptical (in the center of
the Virgo cluster), traced by 275 to 375 globular clusters (GCs), lead to
discrepant results: 
\cite{MGA11} (who also considered stars) 
deduced cored DM (with orbit modeling), while \cite{AERB14} found
cuspy DM (with Jeans modeling).

\section{Methods}
\label{methods}

There are two classes of methods to handle kinematic data of spherical
systems (see the review of \citealp{Courteau+14_chap5}): 1) matching the
line-of-sight (LOS) velocity moments predicted by the Jeans equation with
those measured in bins of projected radii; 2) fitting the full distribution of
observed tracers in projected phase space (PPS), using 6D distribution
functions (DFs) expressed in energy and angular momentum, $f(E,J)$ 
(i.e. orbit modeling). Each has its weaknesses:
The results with moments depend on the choice of
radial bins (see \citealp{RF13}) and do not use the full PPS
information. Orbit modeling currently 
cannot continuously update parameters of the mass profile.

\citet{WLMG09} developed a PPS method where $f(E,J)$ is separable,
taken from \cite{Wojtak+08}, who derived 
the $J$ term of the DF from first principles and 
checked the separable
form of the DF on cluster-mass $\Lambda$CDM halos. 
While this method is ideal for clusters of galaxies, where dissipational effects are
thought to be unimportant, it is not clear how the separability of the
DF will survive the effects of dissipation expected in galaxies. Unfortunately, this method is
slow, as it involves a triple integral to express the
distribution of tracers in PPS in terms of the DF \citep{DM92}.

\cite*{MBB13} developed a very rapid
alternative, called MAMPOSSt, which is as accurate. MAMPOSSt 
assumes a form for the 3D velocity distribution of
tracers (here, Gaussian), which allows one to predict the distribution of tracers
in PPS as a single integral of the LOS velocity distribution function.

\section{Age matching from galaxy mass modeling}

Analyzing, with the DF method of \cite{WLMG09}, 
the kinematics of isolated SDSS galaxies, traced by their
satellites (stacked in 9 bins of host galaxy  stellar mass and color),
\cite{WM13} discovered that the DM halos of red galaxies are 
more concentrated (2$\,\sigma$ level) 
than those of blue galaxies of the same stellar or halo mass.
This is a stark observational  confirmation of \emph{age
  matching} \citep{HW13}: halos that assemble earlier (more concentrated) lead to
galaxies with older stellar populations (redder). Their trends of
halo concentration vs. mass for red and blue galaxies are well matched by
predictions from age matching theory (Watson, private communication).

\section{M87}
\label{M87}

We trace the DM profile of M87 
using GCs. We use the
SLUGGS \citep{Strader+11} database of GCs, removing all objects classified as
UCDs, or transition objects (i.e. only keeping objects with half-mass radii
lower than 5.25 pc, following \citeauthor{Strader+11}), and also removing all velocities measured before
2008 (which we suspect may contain important systematic errors).
This leaves us with 143 red and 322 blue GCs out to 200 kpc. Since the
spectroscopic sample is incomplete, we adopt the
number density profiles based on photometric samples (Table~11 of
\citeauthor{Strader+11}).

We ran MAMPOSSt for several models. Our standard model has a generalized
NFW DM profile (see Sect.~\ref{Fornax}) with the concentration-mass ($c-M$)
relation found in $\Lambda$CDM halos with Planck cosmology \citep{DM14}. Its
stellar mass is fixed to 
$10^{12} M_\odot$ (yielding $M/L_V=8.4$).
Our model also assumes \cite{Tiret+07}
VA with isotropic inner velocities and a transition radius
equal to the effective radius of the tracer. 
\begin{figure}[ht]
\centering
\includegraphics[width=0.49\columnwidth]{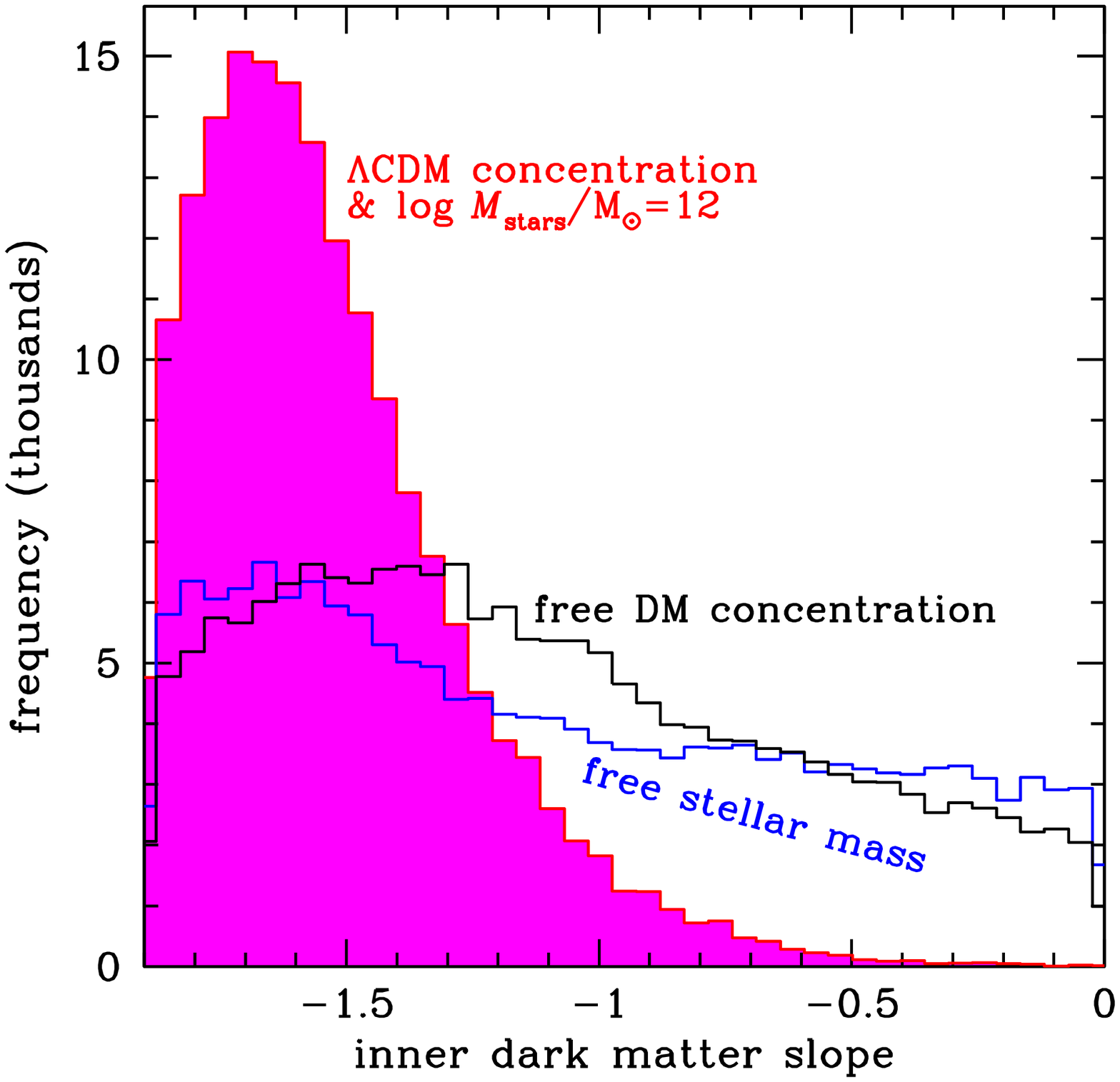}
\includegraphics[width=0.49\columnwidth]{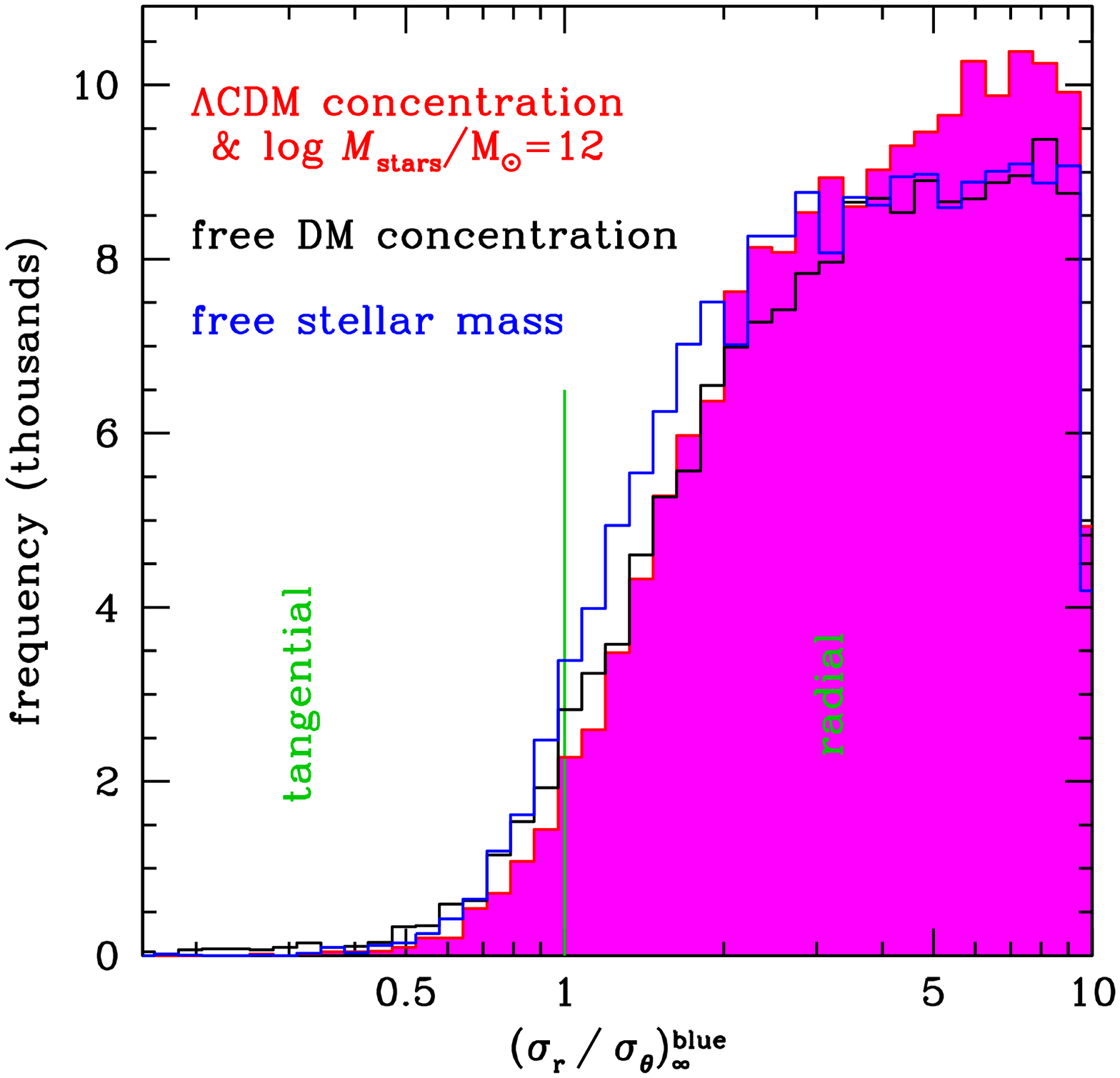}
\caption{Effects of the choice of priors on the PDFs obtained with MAMPOSSt
of the inner DM slope (\emph{left}) and the
outer VA of blue globular clusters 
(\emph{right}) for M87 (traced by red and blue globular clusters).
The \emph{pink shaded regions} correspond to standard prior with $M_{\rm stars}/L_V=8.4$
and the $\Lambda$CDM DM concentration - virial mass relation, while
the other two cases are for free stellar mass (\emph{blue}, best fit corresponding to
$M_{\rm stars}/L_V=25$, which is abnormally high) or free DM concentration
(\emph{black}, best fit at $c=5$, a low value perhaps due to the Virgo cluster).  
}
\label{figM87}
\end{figure}

The left panel of Figure~\ref{figM87} shows how $\gamma$ depends
crucially on the priors of the model. 
While in our standard model ($\Lambda$CDM concentration
and fixed stellar mass, pink shaded region) $\gamma=-1.75\pm0.25$,
significantly \emph{cuspier} than NFW, the
models with either a free concentration (black histogram) or free stellar
mass (blue) allow shallower slopes, including cores, although $\gamma<-1.3$
is again preferred.
Interestingly, the outer VA of the blue GCs is always very
radial in all three cases (right panel of Fig.~\ref{figM87}).
This analysis can be improved by including stellar velocity dispersions
and the surrounding Virgo cluster.

\section{Fornax}
\label{Fornax}

We extracted \cite{WMO09}'s spectroscopic data for the Fornax dSph, only
keeping stars with over 95\% confidence of not being a foreground Milky Way
star (according to the database of \citealp{WMO09}), removing velocity
outliers using the cuts from \cite{Lokas09}, splitting
between $Z$-poor and $Z$-rich components using a critical value for $W'
\equiv \Sigma \rm Mg'$ of 0.5 (i.e., $Z$ rich corresponds to $W' > 0.5$).
This yielded 722 and 1544 $Z$-rich and $Z$-poor stars, respectively.
For both stellar components, 
we assume $n=0.71$ S\'ersic profiles with $R_e = 0.7\,\rm kpc$
\citep{Battaglia+06}.
We ran MAMPOSSt on a standard model with exponentially truncated DM
(as found for tidally truncated subhalos by \citealp{Kazantzidis+04}) 
and free inner DM slope, \cite{Tiret+07} VA, and the
$\Lambda$CDM $c$--$M$ 
relation, allowing the total stellar mass to be free.
\begin{figure}[ht]
\centering
\includegraphics[width=0.49\hsize]{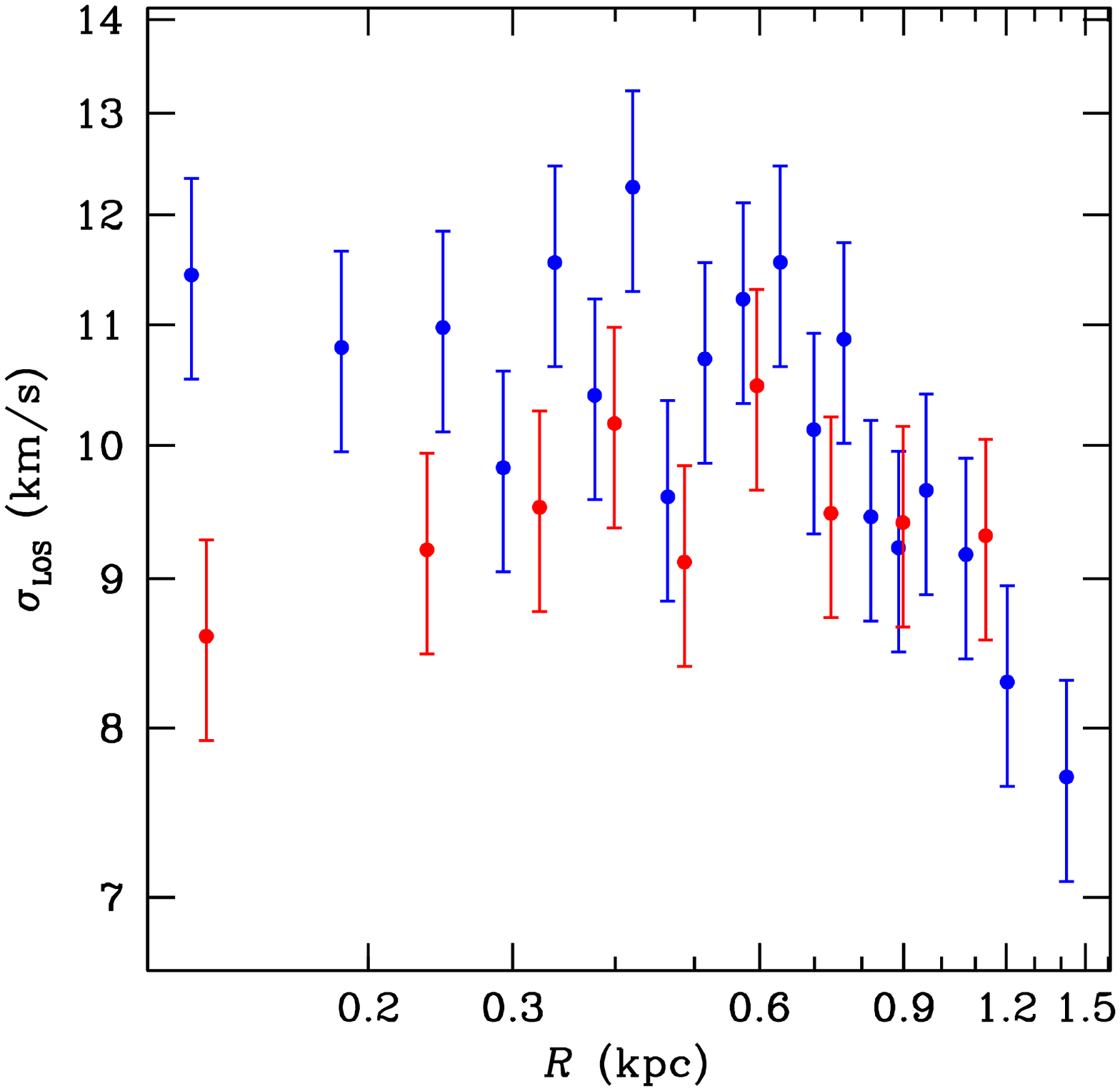} 
\includegraphics[width=0.49\hsize]{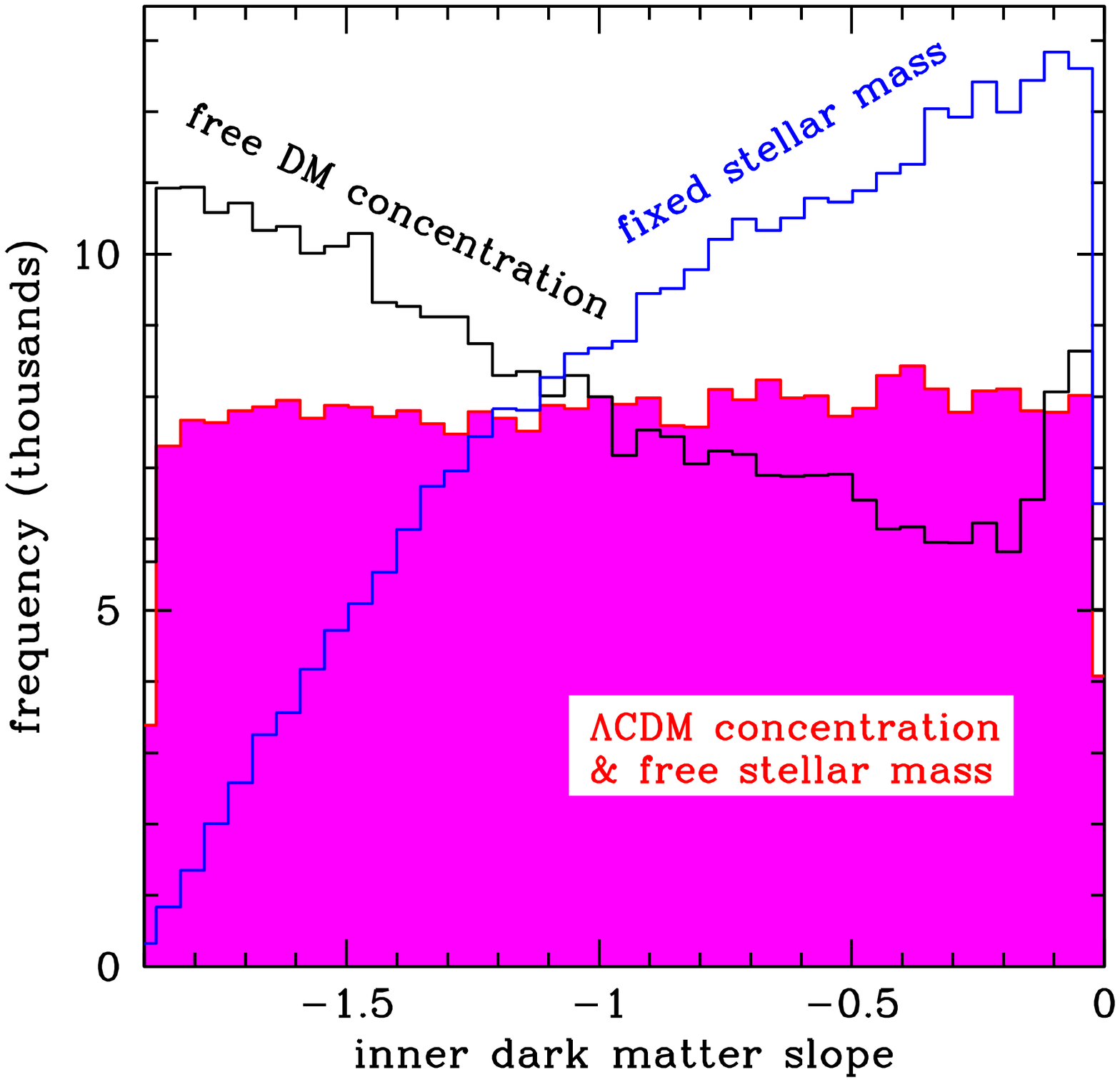} 
\caption{Analysis of the Fornax dSph. \emph{Left}:
Line-of-sight velocity dispersion profiles for the metal-rich
  (\emph{red}) and metal-poor (\emph{blue}) stars.
\emph{Right}: Effects of priors on the PDF of the inner DM slope of Fornax obtained
with MAMPOSSt. 
The \emph{pink shaded region} shows the standard prior with truncated
outer DM, the $\Lambda$CDM DM concentration - virial mass relation, and free
stellar mass, while the other 2 cases are
free DM concentration (\emph{black}, best fit at $c=10$), 
or stellar mass fixed to $10^{7.95} M_\odot$ (\emph{blue}).
}
\label{siglosFornax}
\end{figure}
Figure~\ref{siglosFornax}-left shows that the $Z$-rich population displays lower
velocity dispersions at small projected radii ($R < 0.3\,\rm  kpc$).

Figure~\ref{siglosFornax}-right shows that the inner DM slope is not well
constrained, regardless of the priors on the DM concentration and stellar mass
(despite very different best fits), because if the stellar mass is free, it
dominates the DM within $2\,R_e$ (with asymptotic $\log M_{\rm stars} = 8.2$).
Finally, a robust finding is that the $Z$-rich outer VA is tangential.






\bibliography{master}

\end{document}